\documentclass[twocolumn,secnumarabic,amssymb, nobibnotes, aps, prb]{revtex4-2}
\usepackage{bbm}
\usepackage[latin9]{inputenc}
\usepackage{amsmath}
\usepackage{amssymb}
\usepackage{graphicx}
\usepackage{mathrsfs}
\usepackage{amsfonts}
\usepackage{amsthm}
\usepackage{color}
\usepackage{txfonts}
\usepackage{tabularx}
\usepackage[colorlinks=true,citecolor=blue,linkcolor=blue,urlcolor=blue,anchorcolor=blue]{hyperref}
\hypersetup{colorlinks=true,citecolor=blue,linkcolor=blue,urlcolor=blue}
\begin{document}

\title{Competition of non-Hermitian skin effect and topological localization of corner states observed in circuits}
\author{Chan Tang}
\author{Huanhuan Yang}
\author{Lingling Song}
\author{Xianglong Yao}
\author{Peng Yan}
\author{Yunshan Cao}
\email[Corresponding author: ]{yunshan.cao@uestc.edu.cn}
\affiliation{School of Physics and State Key Laboratory of Electronic Thin Films and Integrated Devices, University of Electronic Science and Technology of China, Chengdu 610054, China}

\begin{abstract}
Exploring topological phases in non-Hermitian systems has attracted significant recent attention. One intriguing question is how topological edge states compete with the non-Hermitian skin effect. Here, we report the experimental observation of corner states in a two-dimensional non-reciprocal rhombus honeycomb electric circuit. We construct non-reciprocal and non-Hermitian circuits by introducing current-direction resolved capacitance between two nodes depends on the current direction. Skin effect thus emerges due to the non-reciprocity and prevails in dragging the corner state into the bulk. The non-Bloch winding number defined in generalized Brillouin zone is adopted to characterize the topological phase transition. Interestingly, we find that the non-Bloch $Z_2$ Berry phase can serve as an invariant to describe the non-Hermitian topology. By tuning the non-reciprocal parameter, we observe unbalanced distribution of corner states emerging on two acute angles of the rhombus lattice, with the localization length of the left corner state increasing exponentially with the degree of non-reciprocity.
\end{abstract}

\maketitle

\section{Introduction}

Topological insulators (TIs) can exhibit edge or surface states following the standard bulk-boundary correspondence \cite{Hasan2010, Qi2011, Chiu2016, Bansil2016, Li2021a}. Topological invariants, such as Zak phase and Chern number, can be defined in terms of Bloch Hamiltonian to dictate the topological phase and phase transition. Higher-order topological insulators (HOTIs), however, allow corner or hinge states with the co-dimension larger than one \cite{Benalcazar2017, Song2017, Schindler2018, Ezawa2018, Xie2018, Ni2019, Li2019, Yang2020, Song2020, Liu2022}. The topological description of HOTIs goes beyond the conventional bulk-boundary correspondence and is depicted by several new topological invariants like $Z_N$ Berry phase \cite{Yang2020} and bulk polarization \cite{Benalcazar2017, Ni2019, Song2020}. According to the Bloch band theory, the energy spectrum obtained from the periodic boundary condition (PBC) should coincide with the one with open boundary condition (OBC), except for the possible in-gap modes.

However, non-Hermitian systems lose the general correspondence of the energy spectrum between PBC and OBC, which leads to the sensitivity of energy band with respect to the open boundaries \cite{Lee2016, Leykam2017, Kunst2018, Herviou2019, Zeng2020, Bergholtz2021, Okuma2020}.
Recently, experimental and theoretical efforts in non-Hermitian topological systems have enabled us to construct the non-Bloch band theory \cite{Yokomizo2019, Song2019, Yang2020a, Helbig2020} by introducing either the non-reciprocal coupling \cite{Lieu2018, Gong2018, Yao2018, Lee2019, Ezawa2019a, Ezawa2019b, Helbig2020, Wu2021, Liu2021} or parity-time symmetry \cite{Weimann2017, Zhang2019, Liu2019, Liu2020, Wu2020, Ezawa2021}. Such non-Hermitian Hamiltonians can be converted into Hermitian ones under a similarity transformation, which is referred to as the pseudo-Hermicity \cite{Fruchart2021}.
To recover the bulk-boundary correspondence, the concept of generalized Brillouin zone (GBZ) has been developed.
Exotic topological invariant, e.g., non-Bloch winding number, is defined in GBZ to discribe topological states for an open system \cite{Lieu2018, Gong2018, Yao2018, Yin2018, Longhi2019, Guo2021}. Experimental evidence of non-Hermitian skin effect in one-dimensional (1D) system has been found to well support the non-Bloch band theory \cite{Weimann2017, Helbig2020, Liu2021}.
Recently, higher-order non-Hermitian skin effects were proposed \cite{Lee2019, Ezawa2019a, Ezawa2019b, Liu2019, Okugawa2020, Kawabata2020, Wu2021}.
However, the experimental observation is still rare, with only few exception \cite{Zou2021}. We are interested in the following issue: How does the localized corner state compete with the non-Hermitian skin effect.

Recently, it is shown that the flexible selection of circuit components provides a convenient way to explore topological physics \cite{Lee2018, Imhof2018}.
Rich topological states have been observed in simple inductor-capacitor (LC) circuits, including topological Chern circuit \cite{Hofmann2019}, HOTIs \cite{Yang2020, Song2020, Liu2022}, weak topological insulators \cite{Yang2022}, square-root higher-order Weyl semimetals \cite{Song2022}, as well as non-Hermitian skin effect \cite{Okuma2020, Helbig2020, Liu2021, Zou2021}.
It was generally believed that topological corner states are hosted for a breathing honeycomb lattice in the absence of non-reciprocity, while non-Hermitian skin effect appears in the absence of breathing but with non-reciprocity. The critical issue we aim to tackle is how these two effects compete with each other when both non-reciprocity and non-Hermiticity are present.
In this work, we show theoretically and experimentally that topological corner states appear at two acute angles in a two-dimensional (2D) non-reciprocal honeycomb electrical circuit and they dramatically evolve with the increasing non-reciprocal strength.
We demonstrate that the localization nature of corner state can be smeared out by the non-Hermitian skin effect such that the localization length increases with the degree of the non-reciprocity.
Besides the non-Bloch winding number, we propose the non-Bloch $Z_2$ Berry phase defined in the GBZ to characterize the non-Hermitian topological phase and phase transition.

This paper is organized as follows: In Sec. \ref{secII}, we construct a rhombus lattice, calculate the admittance spectrum, evaluate the topological invariants and analyze the role of the non-Hermitian skin effect on corner states. In Sec. \ref{secIII}, we experimentally detect the corner states through voltage distribution and measure the unbalanced distribution of two corner states by tuning the non-reciprocity. We discuss and conclude our findings in Sec. \ref{secIV}.

\section{2D non-reciprocal rhombus honeycomb topolectric circuit }\label{secII}

\subsection{Circuit Laplacian}

\begin{figure}
  \centering
  \includegraphics[width=0.95\linewidth]{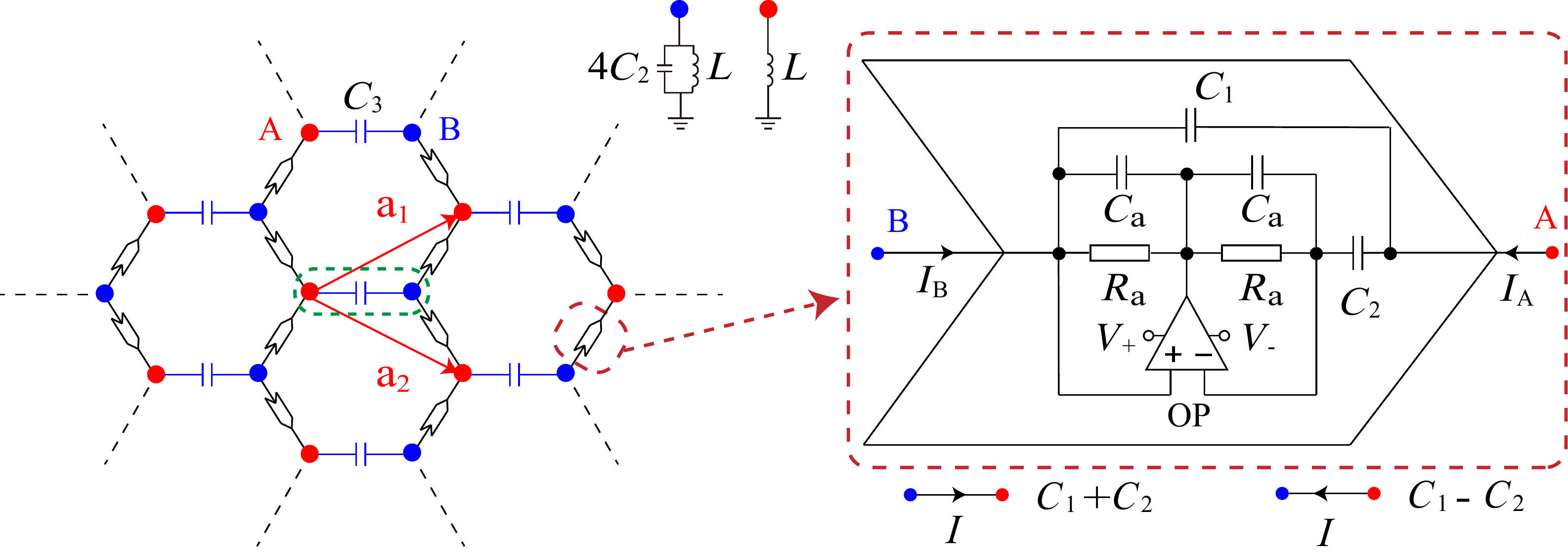}
  \caption{Schematics of the 2D non-reciprocal honeycomb LC circuit with PBC. The dashed green box shows the unit cell consisting of two nodes A (Red) and B (Blue) with the intracell coupling $C_3$ and non-reciprocal intercell coupling $C_1\pm C_2$. The dashed red box shows the operational amplifier (OP) arranged as INIC allowing the non-reciprocal capacitance. }\label{Fig1}
\end{figure}

We first consider a LC network posing honeycomb lattice, as shown in Fig. \ref{Fig1}. Each unit cell includes two nodes labeled by A (red) and B (blue) with reciprocal intracell coupling $C_3$ and non-reciprocal intercell coupling $C_1\pm C_2$. Each node is grounded with a different combination of inductance and capacitance to guarantee the same on-site admittance. All capacitance and induction parameters in our theoretical calculations are real and correspond to the values of the real electrical components. The non-reciprocal capacitance coupling is realized by using operational amplifiers designed as current-inversion negative impedance converters (INIC) (see dashed red box in Fig. \ref{Fig1}), thus we can switch the sign of $C_2$ as reversing the direction of current (see Appendix \ref{appendix:A} for details).

The current and voltage in the electric circuit follow Kirchhoff's law. Considering the monochromatic frequency response, we have
\begin{equation} \label{I}
I_{a}(\omega)=\sum_{b}J_{ab}(\omega)V_{b}(\omega),
\end{equation}
with the circuit Laplacian
\begin{equation} \label{Laplacian}
J_{ab}(\omega)=i\omega\left[-C_{ab}+\delta_{ab}\left(\sum_{n} C_{an}-\frac{1}{\omega^2L_a}\right)\right],
\end{equation}
where $a,b,n=1,2,...$ label different nodes, $I_a$ is the current flowing into node $a$, $V_b$ is the voltage on node $b$, $L_a$ is the grounded inductance at node $a$, and $\delta_{ab}$ represents the Kronecker delta function. The sum is taken over all nearest-neighbour nodes.

\subsection{Periodic boundary conditions}

We choose the two basis vectors of the honeycomb lattice as $\mathbf{a}_1=\frac{\sqrt{3}}{2}\hat{x}+\frac{1}{2}\hat{y}$, $\mathbf{a}_2=\frac{\sqrt{3}}{2}\hat{x}-\frac{1}{2}\hat{y}$ with the unit lattice constant (see Fig. \ref{Fig1}).
We first consider the infinite circuit with PBC. Then, the circuit Laplacian in momentum space can be expressed as a $2\times 2$ matrix

\begin{equation} \label{J}
\mathcal{J}(\omega,{\pmb k})=-i\omega\left(
              \begin{array}{cc}
                h_0 & h_1(\mathbf{k}) \\
                h_2(\mathbf{k}) & h_0 \\
              \end{array}
            \right),
\end{equation}
with
\begin{eqnarray}
   &&h_0=\frac{1}{\omega^2 L}-C_0, ~~~C_0=2C_+ +C_3,\\
   &&h_1(\mathbf{k})=C_3 + 2C_+e^{-i\frac{\sqrt{3}}{2}k_x}\cos\frac{ k_y}{2}, \\
   &&h_2(\mathbf{k})=C_3 + 2C_-e^{i\frac{\sqrt{3}}{2}k_x}\cos\frac{ k_y}{2},~~~C_\pm=C_1\pm C_2 .
\end{eqnarray}
In conventional tight-binding formalism, $h_{1(2)}(\mathbf{k})$ is usually taken as the element of a Bloch Hamiltonian and the Hermicity requires $h_2(\mathbf{k})=h_1^*(\mathbf{k})$.
However, the non-reciprocal coupling leads to the non-Hermicity, i.e., $h_2(\mathbf{k})\neq h_1^*(\mathbf{k})$ in the above equations.
At the resonant frequency $\omega_0=1/\sqrt{L C_0}$, the diagonal elements of circuit Laplacian vanish, i.e., $h_0=0$, the system then has a chiral symmetry $\sigma_z^{-1} \mathcal{J}\sigma_z=-\mathcal{J}$ with $\sigma_z=\text{diag}\{1,-1\}
$ being the Pauli matrix.
By diagonalizing Eq. (\ref{J}), we obtain the following symmetric two-band spectra with respect to the zero-admittance
\begin{equation}\label{twoband}
j_\pm(\mathbf{k})=\pm i\omega_0 \sqrt{h_1(\mathbf{k})h_2(\mathbf{k})}.
\end{equation}
The admittance spectra is purely imaginary for Hermitian system (for Hamiltonian $\mathcal{H}=i\mathcal{J}/\omega$). Due to $h_2(\mathbf{k})\neq h_1^*(\mathbf{k})$, the admittance spectra (\ref{twoband}) takes complex values. However, for a finite system, the purely imaginary spectra emerge (see below), which is inconsistent with spectra (\ref{twoband}) by assuming PBC. Next, we will address this issue.

\subsection{Open boundary conditions}

\begin{figure}[htbp!]
  \centering
  \includegraphics[width=0.95\linewidth]{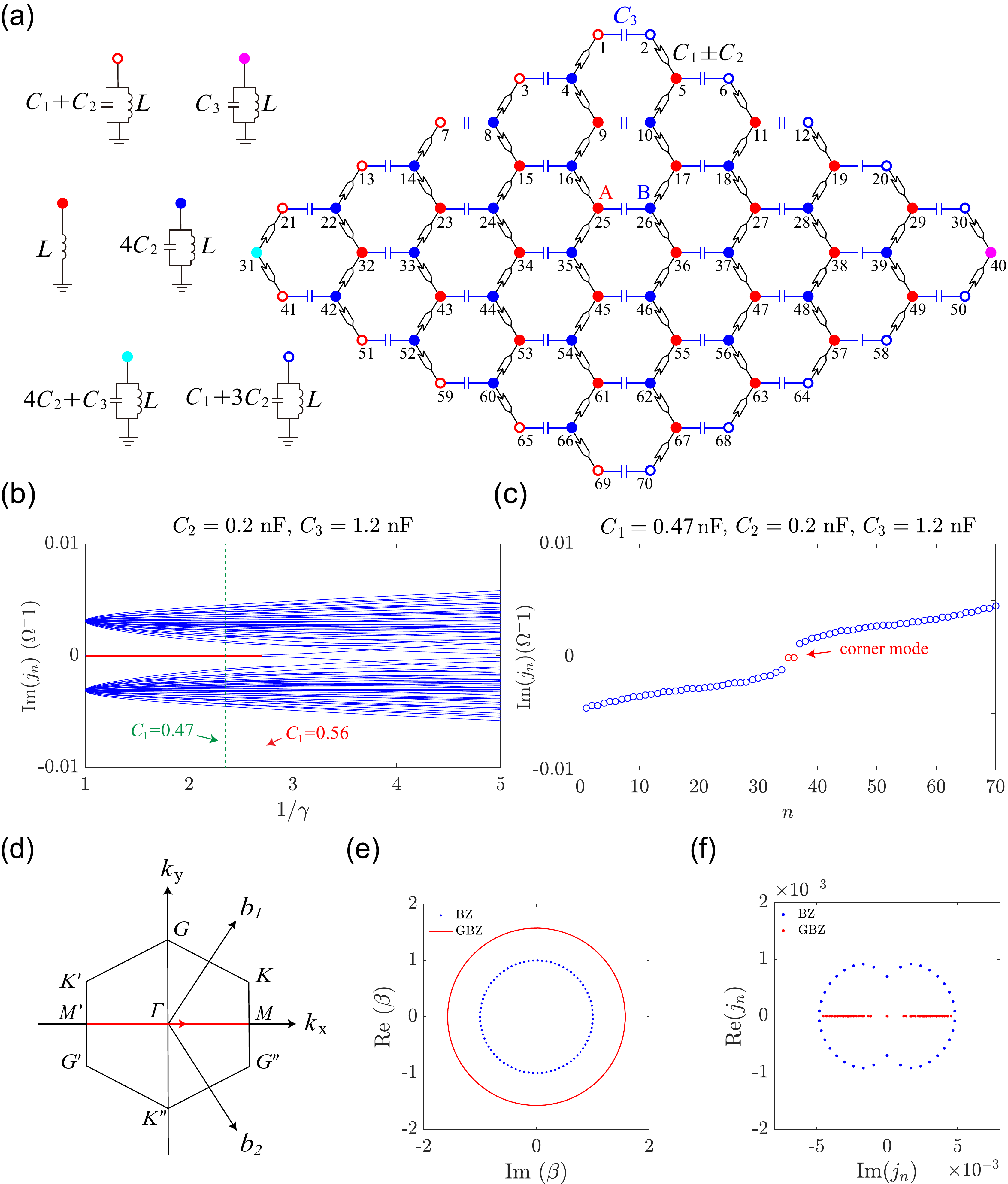}\\
  \caption{(a) A finite honeycomb system with 70 nodes. The red and blue nodes are connected to ground by paralleling with a different combination of inductance and capacitances.
  (b) The admittance band structure evolves with $1/\gamma$, where $C_2=0.2$nF and $C_3=1.2$nF are fixed. Red lines represent the zero mode.
  (c) Admittance spectrum at $C_1=0.47$ nF marked with green line in (b).
  (d) The conventional BZ.
  (e) Red loop stands for GBZ, with $k_x$ ranging from $M'$ to $M$ in BZ. The blue unit circle is defined in the original BZ.
  (f) The complex eigenvalues in BZ and purely imaginary ones in GBZ. }\label{Fig2}
\end{figure}

We design a finite rhombus lattice with $N$ nodes as shown in Fig. \ref{Fig2} (a), with all nodes being labeled. The circuit Laplacian in real space is an $N\times N$ matrix. It reads
\begin{equation}\label{J2}
J_N=-i\omega\left(
              \begin{array}{ccccccc}
              h_0 & C_3 & 0 & C_+ & 0 & 0 & \cdots  \\
              C_3 &  h_0 & 0 & 0 & C_- & 0 & \cdots  \\
               0  &  0  &  h_0 & C_3 & 0 & 0 & \cdots \\
              C_- & 0 & C_3 &  h_0 & 0 & 0 & \cdots  \\
              0 & C_+ & 0 & 0 &  h_0 & C_3 & \cdots  \\
              0 & 0 & 0 & 0 & C_3 &  h_0 & \cdots    \\
              \vdots & \vdots & \vdots & \vdots & \vdots & \vdots & \ddots  \\
              \end{array}
            \right)_{N\times N},
\end{equation}
with $N=70$ in the following calculations and experiments. The eigenequation of Eq. \eqref {J2} can be expressed as
\begin{equation}\label{JNeq}
J_N\psi_n=j_n \psi_n, ~~~n=1,2,...N,
\end{equation}
where $j_n$ and $\psi_n$ are the eigen-admittances and eigenfunctions, respectively.
To facilitate the analysis, we define a nonreciprocal parameter $\gamma=C_2/C_1~(0\leq\gamma<1)$. The reciprocal case corresponds to $\gamma=0$.

We first evaluate the admittance spectrum $j_n$ as a function of $1/\gamma$ at resonant frequency shown in Fig. \ref{Fig2}(b), with fixed capacitors $ C_2=0.2$ nF, $ C_3=1.2$ nF, and inductance $L=75~\mu$H. A representative example of admittance spectrum for $C_1=0.47$ nF is plotted in Fig. \ref{Fig2}(c), in which the two corner states are marked by red circles. We find that eigen-admittances for OBC are purely imaginary, which does not match the complex spectra (\ref{twoband}) obtained by PBC.

To gain more insight, we re-write the admittance \eqref {J2} as $J_N=-i\omega H$. The eigenequation for $H$ is $H\psi_n=E_n \psi_n$, which has the same eigenfunction of $J_N$, with the eigenvalue $E_n=j_n/(-i\omega)$. Since $j_n$ is purely imaginary, $E_n$ is purely real. Then, we take a similarity transformation to the Hamiltonian
\begin{equation}\label{similar}
 S^{-1}HS=\bar{H},
\end{equation}
with $S=\text{diag}\{1,1,r^{-1},r^{-1},r,r...\}$. Here, we choose $r\!=\!\sqrt{C_+/C_-} ~(r\geq 1)$, to ensure the transformed Hamiltonian $\bar{H}$ being Hermitian (see Appendix \ref{appendix:B} for details).
Without loss of generality, we chosse $C_- >0 ~(C_1>C_2)$ to garantee the Hermicity of $\bar{H}$.
Since the similarity transformation does not change the eigenvalues, i.e.,
$\bar{H}\bar{\psi}_n=E_n\bar{\psi}_n$ with $\bar{\psi}_n=S^{-1}\psi_n$, $E_n$ is therefore real which explains the purely imaginary eigenvalue $j_n=-i\omega E_n$ for OBC. We thus call $H$ a pseudo-Hermitian Hamiltonian.

Next, by assuming the PBC, we express $\bar{H}$ in momentum space as
\begin{equation}\label{TransH}
 \bar{\mathcal{H}}(\mathbf{k})=\left(
              \begin{array}{cc}
                0 & q^*(\mathbf{k}) \\
                q(\mathbf{k}) & 0 \\
              \end{array}
            \right),
\end{equation}
with
\begin{equation}
 q(\mathbf{k})=C_3+2\sqrt{C_+C_-}e^{i\frac{\sqrt{3}}{2}k_x}\cos\frac{ k_y}{2}.
\end{equation}
It is noted that Eq. \eqref{TransH} is equivalent to replacing $k_x$ by $k_x-i\frac{2}{\sqrt{3}}\ln r$ in non-Bloch Hamiltonian elements $h_{1(2)}(\mathbf{k})$ in \eqref{J}. As an example, by taking the range of $k_x$ in $[-\frac{2\pi}{\sqrt{3}}, \frac{2\pi}{\sqrt{3}}]$ [from $M'\to\Gamma\to M$ in the original BZ, see Fig. \ref{Fig2} (d)], we introduce a new phase factor $\beta=\exp[i\frac{\sqrt{3}}{2}(k_x-i\frac{2}{\sqrt{3}}\ln r)]=re^{i\frac{\sqrt{3}}{2}k_x}$ shown by the red circle in Fig. \ref{Fig2}(e), which is the GBZ. We observe a radius expansion compared to the original factor $e^{i\frac{\sqrt{3}}{2}k_x}$ (blue dashed unit circle).
One should notice that circuit implementation does not depend on the relative orientation and distance between nodes, because the hopping strength here is independent of the real distance in the circuit, unlike the case in solids. The symmetry determines the type of circuit lattice.  In this regard, the factor $\sqrt{3}/2$ of $k_x$ is not important for the key results.

Moreover, we plot both the eigenvalues of Eq. \eqref{J} in original BZ and those of Eq. \eqref{TransH} in GBZ in Fig. \ref{Fig2}(f). We find that the eigenvalues are complex in the original BZ and purely imaginary in GBZ. The latter one thus repairs the bulk-boundary correspondence in a finite system with OBC.
The concept of GBZ \cite{Yao2018} and pseudo-Hermitian Hamiltonian develops the non-Bloch theory which paves the way to the study of non-Hermitian topology.

\subsection{Non-Bloch topological invariant}

\begin{figure}
  \centering
  \includegraphics[width=0.95\linewidth]{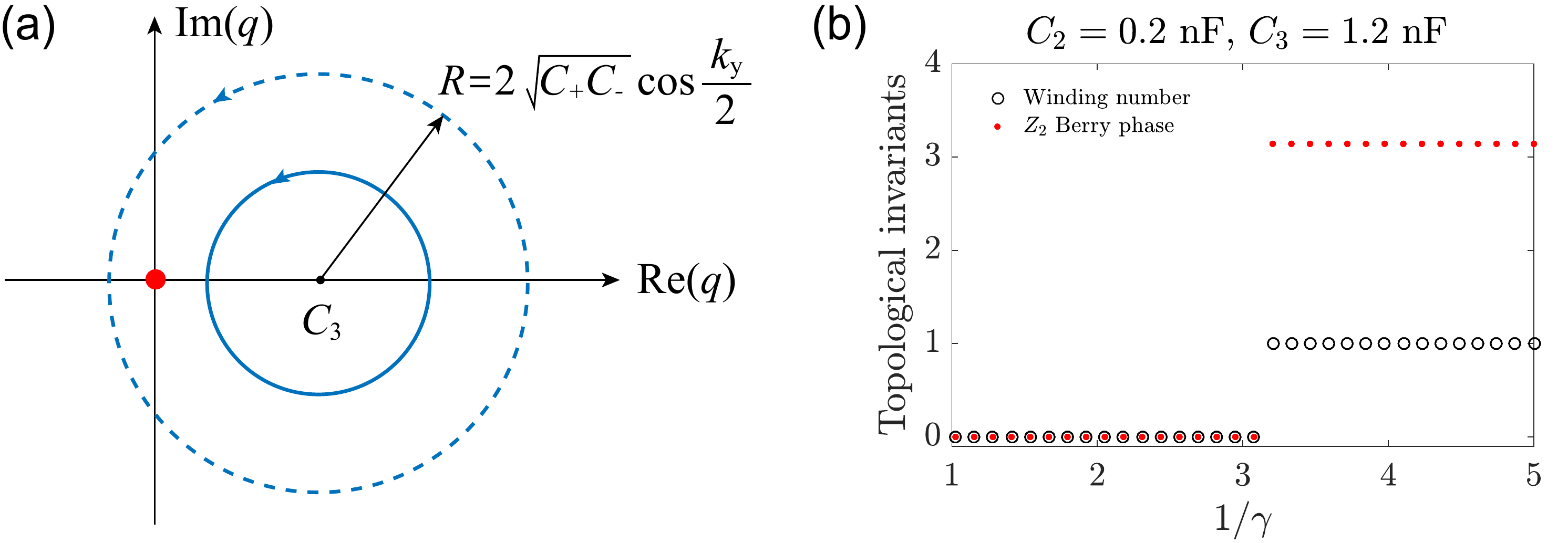}
  \caption{(a) The complex plane of $q(\mathbf{k})$ where the singularity is at the origin. The dashed loop encloses the singularity point while the solid loop does not.
  (b) Topological invariants $W$ and $\theta$ varying with $1/\gamma$ with fixed $C_2=0.2$ nF. The phase transition occurs at $1/\gamma=3.1$, corresponding to $C_1=0.62$ nF. }\label{Fig3}
\end{figure}

\emph{Non-Bloch winding number.} Conventional bulk topological index evaluated from the Bloch Hamiltonian can dictate topological states in the finite system with OBC. For non-reciprocal non-Hermitian systems, one can use the transformed Bloch Hamiltonian to define topological invariants. The so-called non-Bloch winding number is given by \cite{Yao2018}
\begin{equation}\label{W1}
 W=\frac{1}{2\pi i}\oint_{q(\mathbf{k})}\frac{1}{q}dq.
\end{equation}
The integral contour is along $q(\mathbf{k})$ on the complex plane. The value of $W$ depends on if the integral path encloses the singularity point at origin, as shown in Fig. \ref{Fig3}(a). Based on the residue theorem, we obtain
\begin{eqnarray}\label{W3}
&&W=0, ~~\text{for}~~ C_3>2\sqrt{C_+C_-}\left|\cos\frac{ k_y}{2}\right|,\\
&&W=1, ~~\text{for}~~ C_3<2\sqrt{C_+C_-}\left|\cos\frac{ k_y}{2}\right|.
\end{eqnarray}
The topological phase transition occurs at
\begin{equation}\label{W4}
 2\left|\cos\frac{ k_y}{2}\right|=\frac{C_3}{\sqrt{C_1^2-C_2^2}}.
\end{equation}
As an example, we choose $k_y=0$, $C_3=1.2$ nF, and $C_2=0.2$ nF, so the corresponding topological phase transition happens at $C_1=0.62$ nF. The evolution of winding number varying with $1/\gamma$ is shown in Fig. \ref{Fig3}(b), which is close to the phase transition point shown in the admittance structure in Fig. \ref{Fig2}(b) (where the phase transition point approximately locates at $C_1=0.56$ nF ).

\emph{Non-Bloch $Z_2$ Berry phase.} Note that both the original non-Bloch Hamiltonian in Eq. \eqref{J} and the transformed Hamiltonian  Eq. \eqref{TransH} hold the chiral symmetry, i.e., $\sigma_z^{-1} \bar{\mathcal{H}}\sigma_z=-\bar{\mathcal{H}}$, we thus introduce the non-Bloch $Z_2$ Berry phase
\begin{equation}\label{W4}
  \theta=\int_{l} A(\mathbf{k})\cdot d\mathbf{k} ,
\end{equation}
where $A(\mathbf{k})=i\Phi^\dagger(\mathbf{k})\partial_\mathbf{k}\Phi(\mathbf{k})$ is the Berry connection defined though the transformed Hamiltonian \eqref{TransH}, with the lower-band wavefunction $\Phi(\mathbf{k})$. The definition for non-Bloch $Z_2$ Berry phase is generally applicable to all pseudo-Hermitian system.
The integral path $l$ can be chosen as $M'\to \Gamma \to M$, i.e., $k_y=0, k_x \in [-\frac{2\pi}{\sqrt{3}}, \frac{2\pi}{\sqrt{3}}]$. Because of the two-fold rotational symmetry, $\theta$ must be quantized as either $0$ or $\pi$. Figure \ref{Fig3}(b) plots the dependence of $\theta$ as a function of $1/\gamma$ (red dots), and the phase transition point locates at $C_1=0.62$ nF which is identical to the result from the non-Bloch winding number.

\subsection{Non-reciprocal corner states}

\begin{figure}
  \centering
  \includegraphics[width=0.95\linewidth]{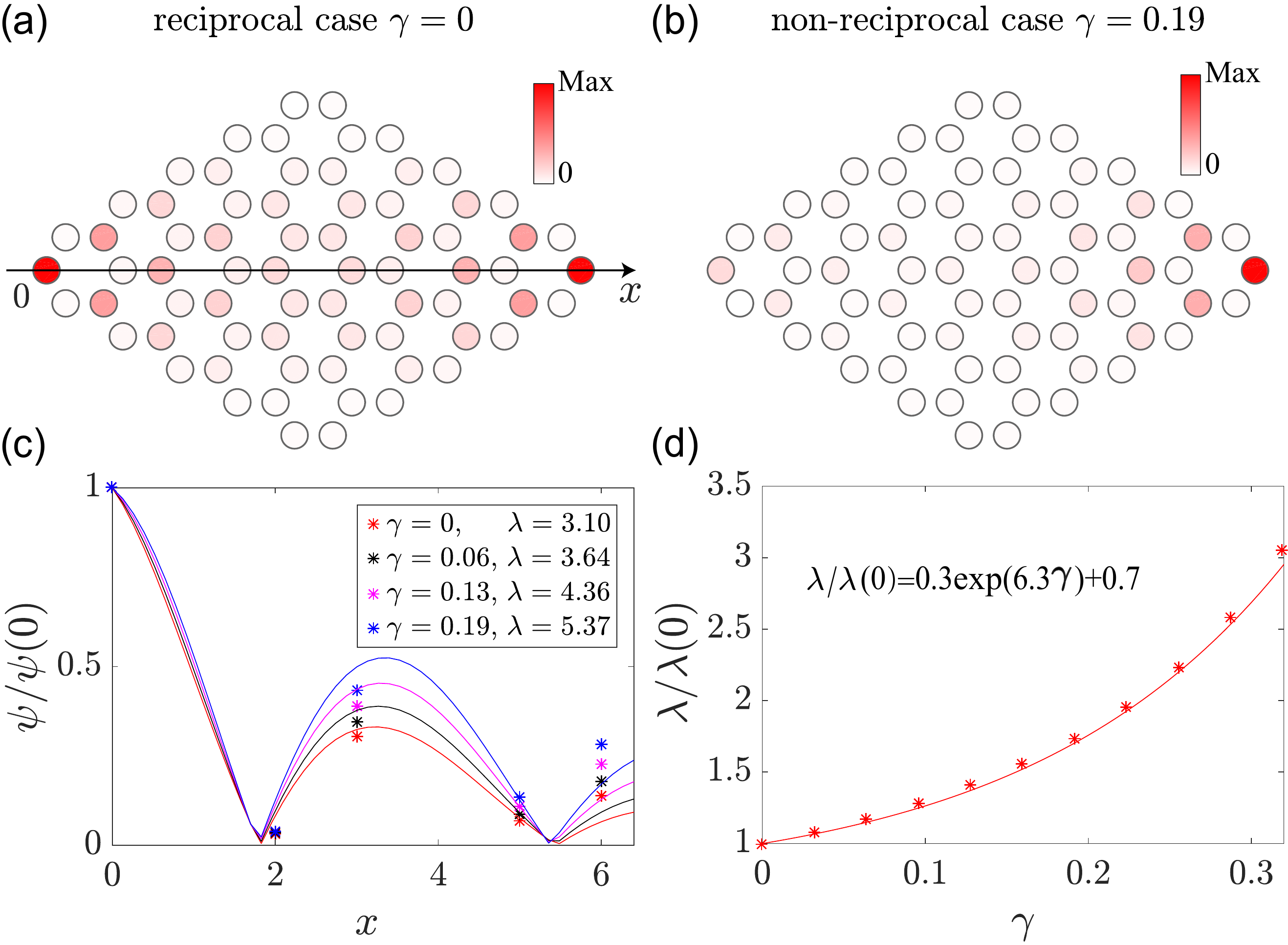}\\
  \caption{Distribution of corner state for reciprocal case $\gamma=0$ (a)
  and non-reciprocal case $\gamma=0.19$ (b). (c) The amplitude of dots on the central line with respect to the dot position. Symbols are numerical calculations and curves are analytical fittings. (d) The localization length (symbols) is fitted exponentially with the non-reciprocal parameter $\gamma$. }\label{Fig4}
\end{figure}

We plot the corner states in reciprocal case $\gamma=0$ and non-reciprocal case $\gamma=0.19$ in Figs. \ref{Fig4}(a) and \ref{Fig4}(b), respectively. The former one shows a symmetric distribution while the latter one demonstrates that the left corner state is smeared out into the bulk due to the non-Hermitian skin effect. To explore how the corner localization competes with the skin effect, we extract the amplitude asymmetry along the central line, see Fig. \ref{Fig4}(c). We find the wavefunction can be well described by formula $|\psi(x)/\psi(0)|=|\cos(\mu  x)|\exp(-x/\lambda)$ with the fitting parameters
$\mu =\{0.846,0.869,0.874,0.877\}$ and $\lambda=\{3.1,3.64,4.36,5.37\}$ for different $\gamma=\{0,0.06,0.13,0.19\}$, respectively. Here, $x$ is the coordinates value of the nodes (corresponding to nodes No. $31\sim35$ in Fig. 2(a)) on the $x$-axis in unit of the lattice constant. The spatial oscillation (a non-zero $\mu$) weakly depends on the circuit non-reciprocity. For the reciprocal case, the topological corner state has a characteristic localization length $\lambda(0)=3.1$ (in unit of the lattice constant). From Fig. \ref{Fig4}(d), we observe that the localization $\lambda$ grows exponentially with $\gamma$. As increasing non-reciprocity, the left corner states survives when $\gamma$ is not so large, with the fitted characteristic non-reciprocal parameter as $\gamma_0=1/6.3=0.16$. We therefore conclude that the non-Hermitian skin effect due to the non-reciprocal coupling tends to drag the topological corner state from the left into the bulk when $\gamma>\gamma_0$. The result does not change significantly with the system size (see Appendix \ref{appendix:C}).

Moreover, the connection between the wave function and node voltage is given by the following formula [6, 15, 26]
\begin{equation}\label{V}
  V_{a}=Z_a I_a=I_a \sum_n\frac{\left|\psi_{n,a}\right|^2}{j_n(\omega)},
\end{equation}
where $n$ runs over all 70 eigenstates, $V_a~(Z_{a})$ represents the voltage (impedance) between the node $a$ and ground. The voltage and impedance would show a resonance peak when one sweeps the frequency. In the following experiments, we measure the voltage response.

\section{Experiments}\label{secIII}

\begin{figure*}[htbp!]
  \centering
  \includegraphics[width=0.95\textwidth]{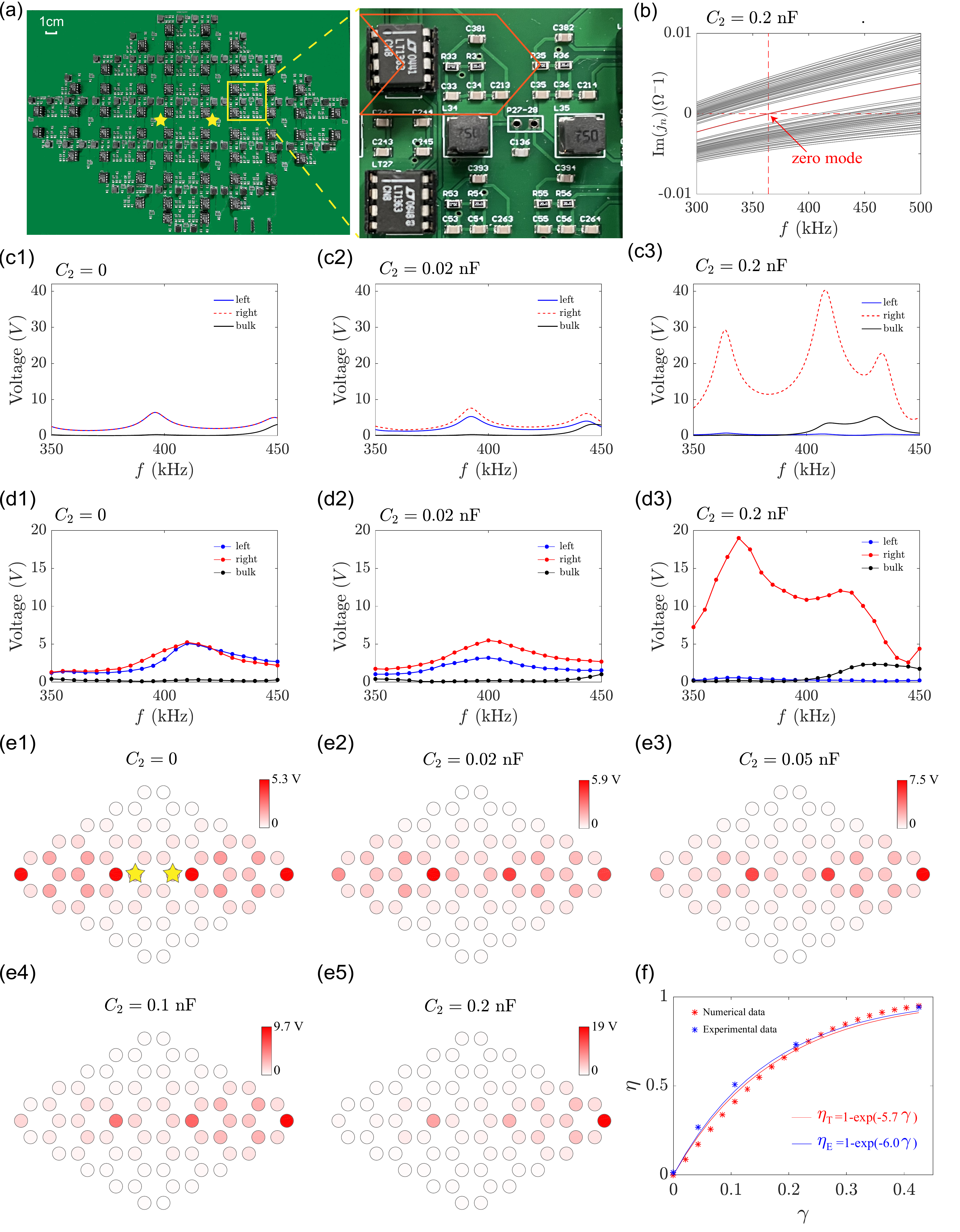}\\
  \caption{(a) The layout of printed circuit board with enlarged partial details in the yellow rectangular box. The orange box represents the components of INIC. Yellow stars indicate the location of the input voltage.
  (b) Calculated spectra of admittance $\mathcal J(\omega)$ varying with the driving frequency. The red solid line in the gap corresponds to the corner states. The intersection of the two red dotted lines represents the corner mode passing through the zero admittance at the resonant frequency. (c1)-(c3) and (d1)-(d3) display the calculated and measured voltage distributions, respectively, with different $C_2=0, 0.02, 0.2$ nF when one sweeps the driving frequency from $ f=350$ to 450 kHz.
  (e1)-(e5) Measured voltage responses at each node for $C_2=0,0.02,0.05,0.1,0.2$ nF at different resonance frequencies $f=410$, 398, 390, 383 and 372 kHz, respectively. The depth of red color represents the magnitude of the voltage. (f) Non-reciprocal ratio $\eta$ varying with $\gamma$. The red asterisks represent numerical calculations, while the red line labels the analytical fitting. Bule asterisks are experimental measurements with the blue line representing the analytical fitting. Other electrical elements are fixed at $C_1=0.47$ nF, $C_3=1.2$ nF, and $L=75~ \mu$H.  }\label{EXP}
\end{figure*}

Figure \ref{EXP}(a) displays the layout of printed circuit board. The parameters are chosen as $C_1=0.47$ nF, $C_3=1.2$ nF, and $L=75~ \mu$H with the quality factor $Q_L=40$. An operational amplifier (LT 1363) with DC voltage $\pm 10$ V is used to realize the INIC (see Appendix \ref{appendix:A} for details).
We measure the voltage response of all nodes by locally inputting sinusoidal driving voltage of $1.5$ V at two nodes marked by yellow stars shown in Fig. \ref{EXP}(a) [see also Fig. \ref{EXP}(e1)].
The zero-admittance mode appears at the resonant frequency as shown in Fig. \ref{EXP}(b).
We numerically calculate the voltage response as a function of driving frequency from $ f=350$ to 450 kHz in Figs. \ref{EXP}(c1)-(c3) with $C_2=0,0.05,0.2$ nF, in which the voltage on the left, right corner, and bulk are marked by blue, red, and black curves, respectively. The secondary peak in Fig. 5(c3) emerges due to the zero-admittance for bulk states when frequency reaches 400 kHz. In the calculations, we take the quality factor into account by adding an imaginary component to the inductance $L\to L(1+i/Q_L)$. Experimental measurements are shown in Figs. \ref{EXP}(d1)-(d3).
The resonance frequencies for voltage peaks of the right corner state are also compared in Table \ref{table1} (Both corner states have the same resonant frequency. However, because the left one is smeared out with increasing $\gamma$, we take the resonant peak of the right one). The slight difference between the numerical results and the experimental measurement is attributed to the quality of capacitances, INIC or device precision. We also demonstrate the measured voltage distributions in Figs. \ref{EXP}(e1)-(e5) on each node for different capacitances $C_2=0,0.02,0.05,0.1,0.2$ nF at the corresponding resonance frequency. As $C_2$ increases, the voltage at the left corner decreases while the right corner increases.
Our experiments confirms the existence of non-Hermitian higher order topological states in 2D non-reciprocal honeycomb, which confirms the theoretical predictions by Ezawa \cite{Ezawa2019b}.

To characterize the non-reciprocity, we define the non-reciprocal ratio $\eta$ as
\begin{equation}\label{eta}
{\eta}={\frac{V_r-V_l}{V_r+V_l}},\\
\end{equation}
where $V_l$ and $V_r$ are the voltage at the left and right corner, respectively.
We evaluate the ratio ${\eta}$ as a function of ${\gamma}$ in Fig. \ref{EXP}(f). It shows an exponential growth by the fitting formula $\eta(\gamma)=1-\exp(-\alpha \gamma)$. The fitting parameter in theoretical calculation (red asterisks) is $\alpha=5.7$ agrees well with the measurement data (blue asterisks) $\alpha=6.0$. The corresponding data of non-reciprocal ratio at chosen $C_2$ are summarized in Table \ref{table1}.

\begin{table}[!htbp]
\centering
\caption{Comparison of theoretical and exprimental results of resonance frequency $f_\mathrm{res}$ and non-reciprocal strength ($\eta_T$ for theory and $\eta_E$ for experiment).}
\renewcommand\arraystretch{1.3}
\begin{tabular}{>{\centering\arraybackslash}p{0.28\linewidth}>{\centering\arraybackslash}p{0.12\linewidth}
>{\centering\arraybackslash}p{0.12\linewidth}>{\centering\arraybackslash}p{0.12\linewidth}
>{\centering\arraybackslash}p{0.12\linewidth}>{\centering\arraybackslash}p{0.12\linewidth}}
  \hline
  \hline
  $C_2$ (nF) & 0 & 0.02 & 0.05 & 0.1 & 0.2 \\
  \hline
  $f_\mathrm{res}$(kHz)~[Theo.] & 397 & 393 & 388 & 380 & 364 \\
  $f_\mathrm{res}$(kHz)~[Exp.] & 410 & 398 & 390 & 383 & 372 \\
  \hline
  $\eta_T$ & 0 & 0.173 & 0.412 & 0.703 & 0.951 \\
  $\eta_E$ & 0.015 & 0.269 & 0.507 & 0.732 & 0.941 \\
  \hline
  \hline
\end{tabular}
\label{table1}
\end{table}

\section{Discussion and conclution}\label{secIV}

Non-reciprocal coupling is one of the main reasons to cause the non-Hermicity. It generates the non-Hermitian skin effect. In our topological circuits, the robustness of corner states is protected by the time-reversal and chiral symmetries. The non-reciprocal coupling would not break the chiral symmetry. However, it breaks the time-reversal symmetry and governs the current flowing preferably in one direction and achieves the one-way signal transmission. Because of the competition between the topological property and the non-Hermitian skin effect, one of the corner states is dragged into the bulk as the non-reciprocal parameter increases. Such non-reciprocal circuits can serve as a good platform to explore novel topological and non-Hermitian physics.

In conclusion, we constructed a non-reciprocal topological circuit with a rhombic honeycomb structure. In addition to the non-Bloch winding number, we propose a non-Bloch $Z_2$ Berry phase to characterize the non-Hermitian topological phase and phase transition. Under open boundary conditions, we numerically calculated the admittance spectrum of the non-reciprocal system and identified the emergence of zero modes pinned to device corners. By measuring the voltage response distribution at resonance frequency, we experimentally observed non-reciprocal corner states on the left and right ends. The non-reciprocal ratio characterizing the relative intensity of wavefunctions of the two corners was found to exponentially increase with the increase of the non-reciprocal parameter, which is fully in line with theoretical calculations. Our finding advances the current understanding of the competition between the non-Hermitian skin effect and localized corner states, and will stimulate the examination of the issue in broad solid-state systems, such as acoustic lattices, photonic crystals, and cold atoms.

\begin{acknowledgments}
\section*{ACKNOWLEDGEMENTS}
This work was supported by the National Key Research Development Program under Contract No. 2022YFA1402802 and the National Natural Science Foundation of China (Grants No. 12074057, No. 11604041, and No. 11704060).
\end{acknowledgments}

\appendix

\section{WORKING PRINCIPLE OF INIC}\label{appendix:A}

\begin{figure}[htbp!]
  \centering
  \includegraphics[width=0.48\textwidth]{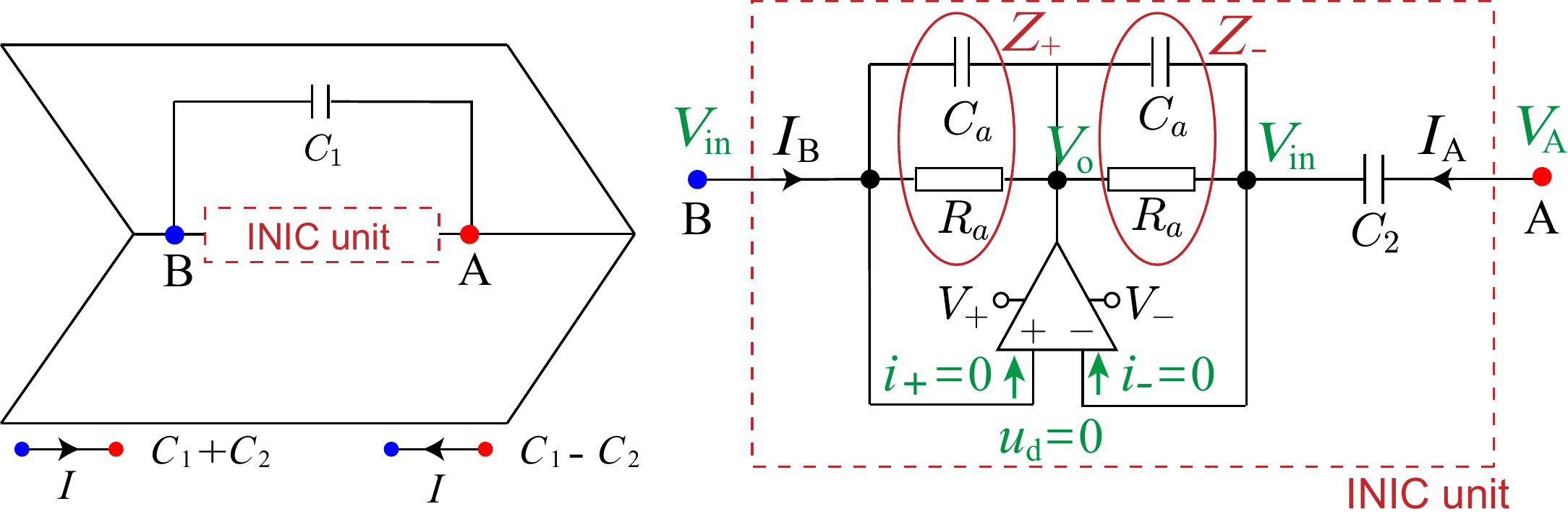}\\
  \caption{The realization of INIC with non-reciprocal intercell couplings $C_1\pm C_2$.}\label{INIC}
\end{figure}

In this Appendix, we provide a detailed analysis for the realization of the non-reciprocal intercell coupling $C_1\pm C_2$, with the circuit details plotted in Fig. \ref{INIC}.

For an operational amplifier, the following three identities hold: $i_+=0$, $i_-=0$, and $u_d=0$. According to the Ohm's law, one can write the following equations for the INIC unit shown in Fig. \ref{INIC} as

\begin{equation} \label{KE}
\begin{aligned}
&I_A=i\omega C_2(V_A-V_{\rm in})=1/Z_-(V_{\rm in}-V_{\rm o}),\\
&I_B=(V_{\rm in}-V_{\rm o})/Z_+,
\end{aligned}
\end{equation}
with $Z_+=Z_-$ being the impedance $1/(i\omega C_a)$ in parallel to $R_a$. From Eqs. \eqref{KE}, we obtain
\begin{equation}
\begin{aligned}
&I_A=i\omega C_2(V_A-V_{\rm in}),\\
&I_B=-i\omega C_2(V_{\rm in}-V_A),\\
\end{aligned}
\end{equation}
which means $I_A=I_B$ ($Z_-/Z_+=1$) instead of $I_A=-I_B$ in a passive circuit element and, as a result, breaks the reciprocity of the circuit. Therefore, the INIC unit can be viewed as $C_2$ ($-C_2$) when current flows into node A (B). Subsquently, one can realize non-reciprocal intercell coupling $C_1\pm C_2$ by paralleling the INIC unit with the capacitor $C_1$.

In our experiments, we use the DC power supply (IT6332A) to provide $V_\pm=\pm10$ V voltage to the operational amplifiers. To avoid the self-excitation of operational amplifier, two capacitors (0.1 $\mu$F and 100 $\mu$F) are paralleled between the output terminals of DC power supplier and the ground, which can filter out the AC signals from the DC voltage sources.

\section{SIMILARITY TRANSFORMATION}\label{appendix:B}
\begin{figure}[htbp!]
  \centering
  \includegraphics[width=0.48\textwidth]{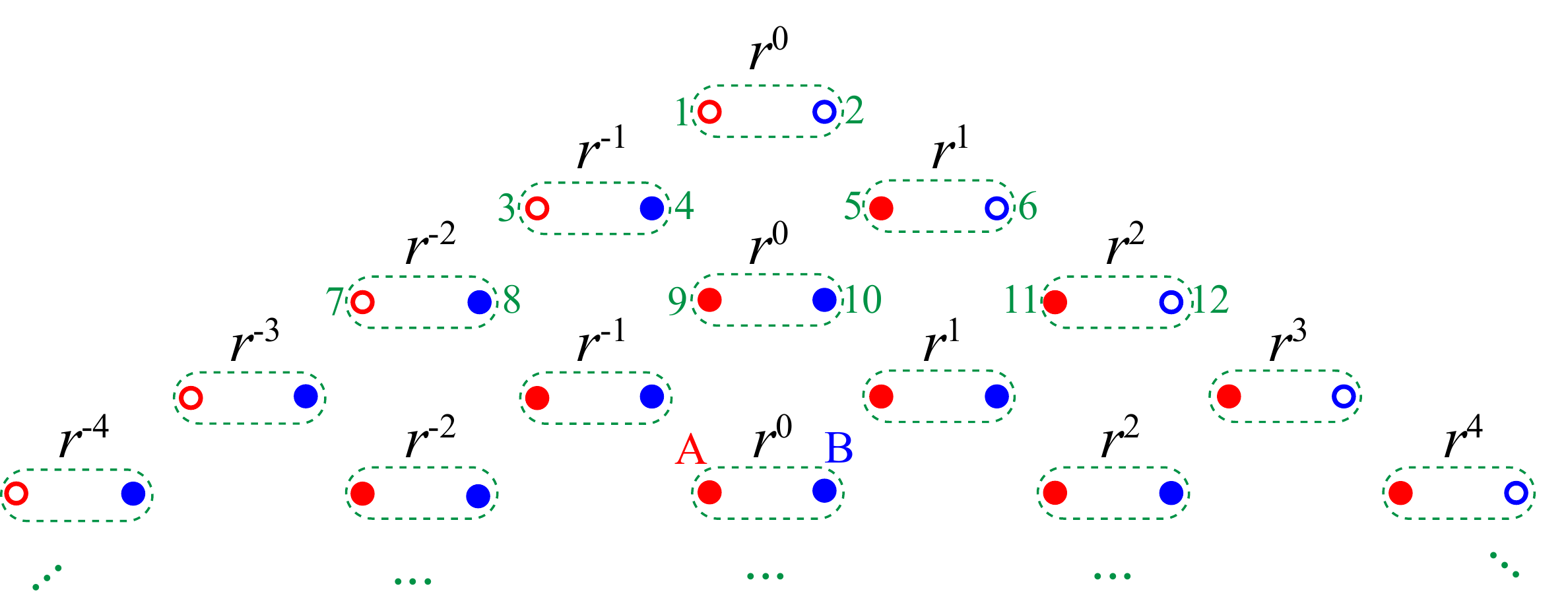}\\
  \caption{Graphic representation of the elements of the diagonal matrix $S$.}\label{Smatrix}
\end{figure}

The similarity transformation matrix in Eq. \eqref{similar} is a diagonal matrix $S=\text{diag}\{1,1,r^{-1},r^{-1},r,r,....\}$. All indexes with labelled numbers are shown in Fig. \ref{Smatrix}. The transformed Hamiltonian reads

\begin{equation}
\bar{H}=\left(
              \begin{array}{ccccccc}
              0 & C_3 & 0 & \sqrt{C_+C_-} & 0 & 0 & \cdots  \\
              C_3 & 0 & 0 & 0 & \sqrt{C_+C_-} & 0 & \cdots  \\
               0  &  0  & 0 & C_3 & 0 & 0 & \cdots \\
              \sqrt{C_+C_-} & 0 & C_3 & 0 & 0 & 0 & \cdots  \\
              0 & \sqrt{C_+C_-} & 0 & 0 & 0 & C_3 & \cdots  \\
              0 & 0 & 0 & 0 & C_3 & 0 & \cdots    \\
              \vdots & \vdots & \vdots & \vdots & \vdots & \vdots & \ddots  \\
              \end{array}
            \right)_{N\times N}.
\end{equation}

\section{SIZE-DEPENDENT EFFECT}\label{appendix:C}
\begin{figure*}[htbp!]
  \centering
  \includegraphics[width=0.95\textwidth]{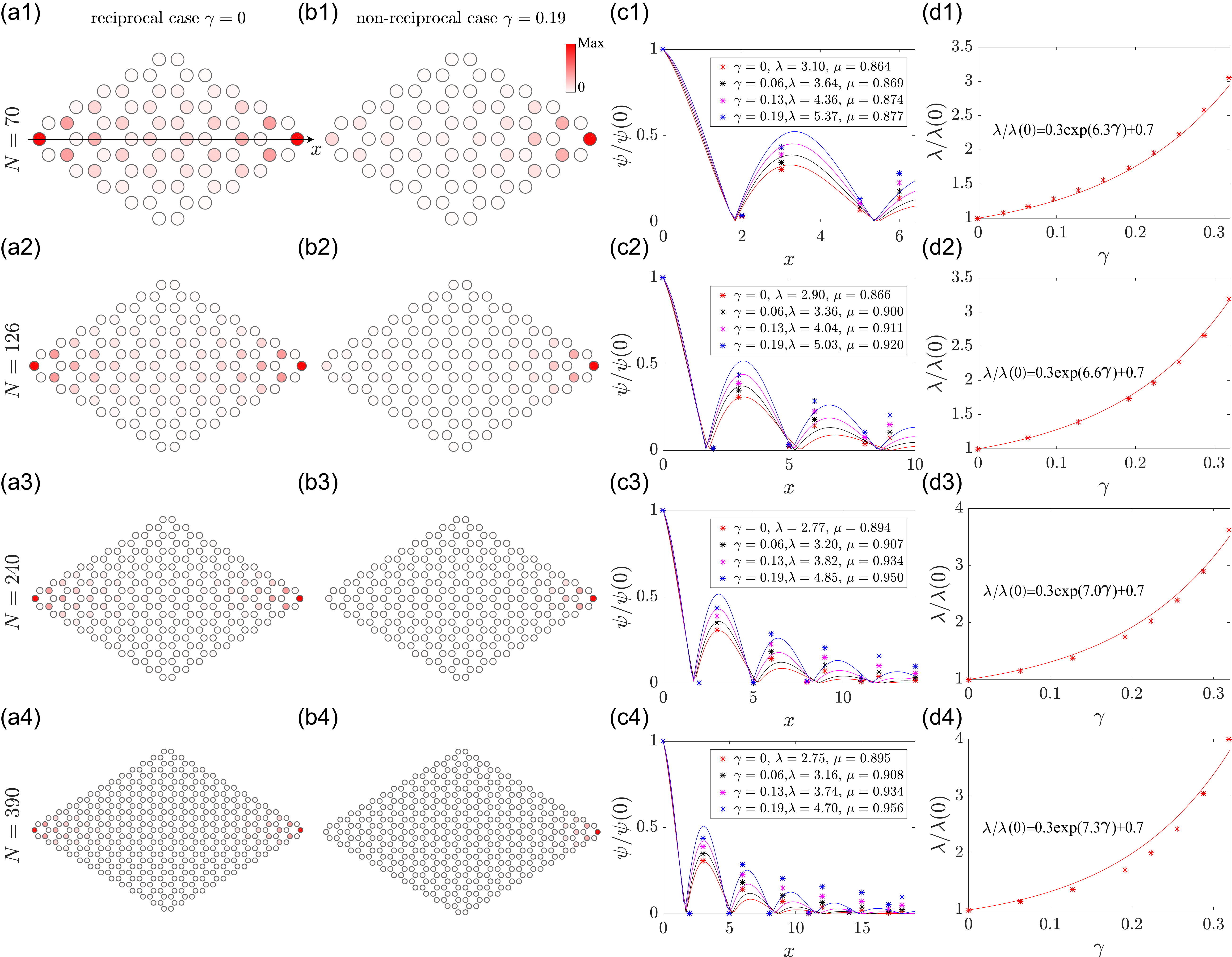}\\
  \caption{Size-dependent corner states distribution in reciprocal (a1-a4) and non-reciprocal cases (b1-b4). The wave function distribution (c1-c4) with respect to the nodes from left end to the bulk center on the x-axis. Symbols are numerical calculations and curves are analytical fittings. The localization length (symbols) is fitted exponentially growing with the non-reciprocal parameter $\gamma$, the characterized non-reciprocal parameter $\gamma_0=1/6.3, 1/6.6, 1/7.0, 1/7.3$ is slightly decreasing when the system becomes larger.}\label{size}
\end{figure*}

A size-dependent zero mode was reported for a special coupled 1-d system \cite{Islam2022}. In this Appendix, we show that our findings is only weakly dependent on the system size. The corner state still survives with the increasing non-reciprocity when the system becomes larger. The non-Hermitian honeycomb lattice hosts corner states which are not affected by the size significantly, see Fig. \ref{size}. By increasing the system size to $N=126, 240, 390$, the corner states are emerged in reciprocal case and smeared out into the bulk in non-reciprocal case. The fitting results show less dependence of $\mu$ on the non-reciprocal parameter. As the system size becoming larger, the fitting value of $\mu$ change slightly with the system size. The fitted localization length exponentially increases with the non-reciprocal parameter in all system sizes. And the characterized non-reciprocal parameter $\gamma_0=1/6.3, 1/6.6, 1/7.0, 1/7.3$ is slightly decreasing when the system becomes larger.

{

\end{document}